\documentclass[aps,twocolumn]{revtex4}

\usepackage{amsmath}
\usepackage{amssymb}
\usepackage{amsfonts}
\usepackage{amsthm}
\usepackage{bm}
\usepackage{color}
\usepackage{mathrsfs}
\usepackage{graphicx}
\usepackage{comment}

\newcommand{\e}{\mathrm{e}}
\renewcommand{\d}{\mathrm{d}}
\newcommand{\voter}{\mathrm{voter}}

\renewcommand{\O}{\mathcal{O}}

\begin{document}

\title{
    Approximate master equations for the spatial public goods game
}
\author{Yu Takiguchi}
\author{Koji Nemoto}
\affiliation{Division of Physics, Hokkaido University, Sapporo 060-0810, Japan}

\begin{abstract}
The spatial public goods game has been used to examine factors that promote cooperation.
Owing to the complexity of the dynamics of this game, previous studies on this model neglected analytical approaches 
and relied entirely on numerical calculations using the Monte Carlo (MC) simulations.
In this paper, we present the approximate master equations (AMEs) for this model.
We report that the results obtained by the AMEs are mostly qualitatively consistent with those obtained by the MC simulations.
Furthermore, we show that it is possible to obtain phase boundaries analytically in certain parameter regions.
In the region where the noise in strategy decisions is very large, the phase boundary can be obtained analytically 
by considering perturbations from the steady state of the voter model.
In the noiseless region, discontinuous phase transitions occur because of the characteristics of the function that represents strategy updating.
Our approach is useful for clarifying the details of the mechanisms that promote cooperation
and can be easily applied to other group interaction models.
\end{abstract}

\maketitle

\section{Introduction}

The selfish behavior of individuals often leads to undesirable outcomes.
Public goods are prone to exploitation by free riders who reap benefits without incurring costs.
This phenomenon is known as a social dilemma or the "tragedy of the commons"\cite{hardin1968tragedy}. 
There is considerable interest in understanding what promotes cooperation, and game theory has been used to delve into this question\cite{nowak2006five}.
Since network reciprocity has been identified as a mechanism that promotes cooperation\cite{nowak1992evolutionary}, 
many studies have examined the effect of network structure on cooperation\cite{hauert2005game,szabo2005phase,vukov2006cooperation,fu2007evolutionary,gomez2007dynamical,assenza2008enhancement,ohtsuki2006simple,santos2008social}.
Among others, pairwise games (e.g., prisoner's dilemma games) on lattices have been studied by numerical calculations using the pair approximation\cite{hauert2005game} or approximations considering larger blocks of nodes\cite{szabo2005phase,vukov2006cooperation} as well as Monte Carlo (MC) simulations.

The spatial public goods game is a fundamental model for the cooperative interactions that occur in groups on a social network\cite{santos2008social,szolnoki2009topology,javarone2016role,perc2017statistical,perc2013evolutionary,battiston2020networks}.
The game has also been studied as a statistical physics model because of its similarity to the spin model.
In previous studies\cite{szolnoki2009topology,javarone2016role}, phase diagrams for this model were derived using MC simulations.
They have revealed that cooperators can survive under conditions where they die out in a well-mixed population,
and that cooperators and defectors can coexist, which is impossible in a well-mixed population.
However, the previous studies only involved numerical calculations because of the complexity of the game dynamics,
and hence, the exact values of the transition points where the cooperators or the defectors become extinct are not known, and an analytical understanding is insufficient.

In this paper, we present the approximate master equations (AMEs)\cite{gleeson2013binary,gleeson2011high,wang2023high} for the spatial public goods game.
This method is an extension of the mean-field and the pair approximation and is more accurate than these.
In addition, the AMEs can be applied to networks other than lattices.
There are two advantages to using this equation. 
First, the AMEs can be a platform for numerical calculation, and also be used to devise analytical approaches for certain parameters.
This makes it possible to reveal the details of the mechanisms that promote cooperation.
Second, the AMEs describe the dynamics of a network of infinite size, enabling us to study the properties of this model in the thermodynamic limit.

\section{Model}

\subsection{Spatial public goods game}

First we explain the public goods game.
We consider a group of $G$ agents, each adopting one of the following strategies: cooperation (C) or defection (D).
A cooperator pays a cost, $1$, and contributes to the public goods.
A defector incurs no cost and does not contribute.
The public goods is denoted as $F(N_{\mathrm{C}})$, a function of the number of cooperators $N_{\mathrm{C}}$, 
and is divided equally among all members of the group.
In accordance with previous studies, we adopt the simplest one: $F(N_{\mathrm{C}}) = r N_{\mathrm{C}}$, where the parameter $r$ is the synergy factor of cooperation.
Therefore, the payoff of a cooperator is $\Pi_{\mathrm{C}} = r N_{\mathrm{C}} / G - 1$, while that of a defector is $\Pi_{\mathrm{D}} = r N_{\mathrm{C}} / G$.

Next, we introduce a network structure into this model.
We consider a $k$-regular random graph with $N$ nodes.
Each node represents an agent who plays the public goods game in a group consisting of $G=k+1$ nodes: itself and its $k$ neighbors.
We adopt the following imitation dynamics as the rule for strategy updating.
A node $x$ and its adjacent role model $y$ are randomly chosen.
The node $x$ imitates the strategy of the role model $y$ with a probability given by the sigmoid function:
\begin{align}
    f(\Pi_x - \Pi_y)
    &\equiv \frac{1}{1 + \e^{(\Pi_x - \Pi_y) / K}},
    \label{imitation}
\end{align}
where $\Pi_i$ denotes the payoff of node $i \ (\in\{x,y\})$, 
and $K$ denotes the uncertainty in the strategy adoptions (or noise).
In the limit $K \to 0$, the node $x$ copies the strategy of the node $y$ if and only if $\Pi_x < \Pi_y$.
Conversely, in the limit $K \to \infty$, the decision on whether to imitate the strategy of the role model is randomly determined regardless of their payoffs.

Initially, cooperators and defectors are distributed uniformly at random.
The strategy is repeatedly updated to obtain the fraction of cooperators, $\rho^{\mathrm{C}}$, in the steady state.
Note that $\rho^{\mathrm{C}}=0$ and $1$ are absorbing states since the strategy of a node changes only if its strategy and the strategy of a role model are different.
The operating parameters are $r$ and $K$.
We define the region where $\rho^{\mathrm{C}}=1$ in the steady state as the C phase, 
the region where $\rho^{\mathrm{C}}=0$ as the D phase, 
and the region where $0<\rho^{\mathrm{C}}<1$ as the (C+D) phase.

\subsection{Mean-field approximation}

Before deriving the AMEs, we consider a differential equation using the mean-field approximation.
Since this approximation cannot represent the network structure, 
it is equivalent to the dynamics in a well-mixed population.
Let $\rho^{\mathrm{C}}(t)$ and $\rho^{\mathrm{D}}(t) \ (= 1 - \rho^{\mathrm{C}}(t))$ denote the fractions of cooperators and defectors at time $t$, respectively.
Here, $\rho^{\mathrm{C}}$ increases when a D-node is chosen as the node updating its strategy and a C-node is chosen as the role model
and the D-node imitates the strategy of the C-node.
This probability is $\rho^{\mathrm{D}} \rho^{\mathrm{C}} f(\Pi_{\mathrm{D}} - \Pi_{\mathrm{C}})$.
Note that the strategies of $(k-1)$ nodes in the group to which the D-node (or the C-node) belongs have not been determined, that is, $N_{\mathrm{C}} = (k-1)\rho^{\mathrm{C}} + 1$.
The decreasing case is also derived similarly.
Thus, the time derivative of $\rho^{\mathrm{C}}$ can be expressed as follows:
\begin{align}
    \frac{\d}{\d t} \rho^{\mathrm{C}}
    &= 
    \rho^{\mathrm{C}} \left( 1 - \rho^{\mathrm{C}} \right) 
    \left[ f(\Pi_{\mathrm{D}} - \Pi_{\mathrm{C}}) - f(\Pi_{\mathrm{C}} - \Pi_{\mathrm{D}}) \right]\nonumber\\
    &=
    \rho^{\mathrm{C}} \left( 1 - \rho^{\mathrm{C}} \right)  \tanh \left( \frac{\Pi_{\mathrm{C}} - \Pi_{\mathrm{D}}}{2K} \right),
    \label{MF}
\end{align}
where
\begin{align}
    \Pi_{\mathrm{C}} = r \frac{(k-1)\rho^{\mathrm{C}} + 1}{k+1} - 1,&&
    \Pi_{\mathrm{D}} = r \frac{(k-1)\rho^{\mathrm{C}} + 1}{k+1}.
\end{align}
Since $\Pi_{\mathrm{C}} < \Pi_{\mathrm{D}}$, $\rho^{\mathrm{C}}$ consistently decreases and eventually reaches zero.
In other words, cooperators cannot survive in a well-mixed population.

\section{Results}

\subsection{Approximate master equation}

We derive the AMEs for this model.
We call a node with strategy $s(\in \{\mathrm{C},\mathrm{D}\})$ that is adjacent to $m(=0,1,\ldots,k)$ cooperators as an $s_m$ node.
The payoff of an $s_m$ node is denoted as $\Pi_{s_m}$ and can be expressed as follows:
\begin{align}
    \Pi_{\mathrm{C}_m} = r \frac{m + 1}{k+1} - 1,&& 
    \Pi_{\mathrm{D}_m} = r \frac{m}{k+1}.
\end{align}
We define the fraction of $s_m$ nodes at time $t$ as $\rho^{s}_m(t)$.
These variables satisfy the following normalization:
\begin{align}
    \sum_{s\in \{\mathrm{C},\mathrm{D}\}}\sum_{m=0}^{k} \rho^s_m = \rho^{\mathrm{C}} + \rho^{\mathrm{D}} = 1.
\end{align}
Further, note that from the two counting methods of the number of CD edges, the following relationship holds:
\begin{align}
    \sum_{m} (k-m) \rho^{\mathrm{C}}_m = \sum_{m} m \rho^{\mathrm{D}}_m.
    \label{CD_edge}
\end{align}
The AMEs are differential equations in $\rho^{s}_m$ given as follows (see \cite{gleeson2013binary} and appendix \ref{deviation of AME} for details):
\begin{align}
    \frac{\d}{\d t} \rho^{\mathrm{C}}_m
    =& 
    - W^{\mathrm{C} \to \mathrm{D}}_m \rho^{\mathrm{C}}_m
    + W^{\mathrm{D} \to \mathrm{C}}_m \rho^{\mathrm{D}}_m\nonumber\\
    &
    - \beta^{\mathrm{C}} (k - m) \rho^{\mathrm{C}}_m
    + \beta^{\mathrm{C}} (k - m + 1) \rho^{\mathrm{C}}_{m-1}\nonumber\\
    &
    - \gamma^{\mathrm{C}} m \rho^{\mathrm{C}}_m
    + \gamma^{\mathrm{C}} (m+1) \rho^{\mathrm{C}}_{m+1},\nonumber\\
    \frac{\d}{\d t} \rho^{\mathrm{D}}_m
    =& 
    - W^{\mathrm{D} \to \mathrm{C}}_m \rho^{\mathrm{D}}_m
    + W^{\mathrm{C} \to \mathrm{D}}_m \rho^{\mathrm{C}}_m\nonumber\\
    &
    - \beta^{\mathrm{D}} (k - m) \rho^{\mathrm{D}}_m
    + \beta^{\mathrm{D}} (k - m + 1) \rho^{\mathrm{D}}_{m-1}\nonumber\\
    &
    - \gamma^{\mathrm{D}} m \rho^{\mathrm{D}}_m
    + \gamma^{\mathrm{D}} (m+1) \rho^{\mathrm{D}}_{m+1},\label{AME}
\end{align}
where we define $\rho^{\mathrm{C}}_{-1}=\rho^{\mathrm{C}}_{k+1}=\rho^{\mathrm{D}}_{-1}=\rho^{\mathrm{D}}_{k+1}=0$,
and $W^{s\to s'}_{m}$ denotes the transition probability from $s_m$ to $s'_m$.
\begin{align}
    W^{\mathrm{C} \to \mathrm{D}}_m
    &\equiv \frac{k - m}{k} \left\langle f(\Pi_{\mathrm{C}_m} - \Pi_{\mathrm{D}}) \right\rangle,\nonumber\\
    W^{\mathrm{D} \to \mathrm{C}}_m
    &\equiv \frac{m}{k} \left\langle f(\Pi_{\mathrm{D}_m} - \Pi_{\mathrm{C}}) \right\rangle,\\
    \beta^{\mathrm{C}}
    &\equiv \frac{\sum_{m'} m' W^{\mathrm{D} \to \mathrm{C}}_{m'} \rho^{\mathrm{D}}_{m'}}{\sum_{m'} m' \rho^{\mathrm{D}}_{m'}},\nonumber\\
    \gamma^{\mathrm{C}}
    &\equiv \frac{\sum_{m'} m' W^{\mathrm{C} \to \mathrm{D}}_{m'} \rho^{\mathrm{C}}_{m'}}{\sum_{m'} m' \rho^{\mathrm{C}}_{m'}},\nonumber\\
    \beta^{\mathrm{D}}
    &\equiv \frac{\sum_{m'} (k-m') W^{\mathrm{D} \to \mathrm{C}}_{m'} \rho^{\mathrm{D}}_{m'}}{\sum_{m'} (k-m') \rho^{\mathrm{D}}_{m'}},\nonumber\\
    \gamma^{\mathrm{D}}
    &\equiv \frac{\sum_{m'} (k-m') W^{\mathrm{C} \to \mathrm{D}}_{m'} \rho^{\mathrm{C}}_{m'}}{\sum_{m'} (k-m') \rho^{\mathrm{C}}_{m'}},
\end{align}
where we define
\begin{align}
    &\left\langle f(\Pi_{\mathrm{C}_m} - \Pi_{\mathrm{D}}) \right\rangle\nonumber\\
    &\equiv 
    \sum_{m'=0}^k 
    \frac{m' \rho^{\mathrm{D}}_{m'}}{\sum_{m''=0}^k m'' \rho^{\mathrm{D}}_{m''}}
    f(\Pi_{\mathrm{C}_m} - \Pi_{\mathrm{D}_{m'}}),\nonumber\\
    &\left\langle f(\Pi_{\mathrm{D}_m} - \Pi_{\mathrm{C}}) \right\rangle\nonumber\\
    &\equiv
    \sum_{m'=0}^k 
    \frac{(k - m') \rho^{\mathrm{C}}_{m'}}{\sum_{m''=0}^k (k - m'') \rho^{\mathrm{C}}_{m''}}
    f(\Pi_{\mathrm{D}_m} - \Pi_{\mathrm{C}_{m'}}).
\end{align}
Notice that the AMEs do not account for the loop structure, and therefore, it corresponds to the dynamics on the Bethe lattice.
We begin with an initial state where cooperators and defectors are randomly distributed.
\begin{align}
    \rho^{\mathrm{C}}(0) &= \frac{1}{2},\nonumber\\
    \rho^{\mathrm{C}}_m(0) 
    &= 
    \rho^{\mathrm{C}}(0)
    \begin{pmatrix}
        k\\
        m
    \end{pmatrix}
    \left( \rho^{\mathrm{C}}(0) \right)^m
    \left( 1 - \rho^{\mathrm{C}}(0) \right)^{k-m},\nonumber\\
    \rho^{\mathrm{D}}_m(0) 
    &= 
    \left( 1 - \rho^{\mathrm{C}}(0) \right)
    \begin{pmatrix}
        k\\
        m
    \end{pmatrix}
    \left( \rho^{\mathrm{C}}(0) \right)^m
    \left( 1 - \rho^{\mathrm{C}}(0) \right)^{k-m}.
\end{align}
By solving the AMEs numerically, we obtain $\rho^{\mathrm{C}}(t) = \sum_{m} \rho^{\mathrm{C}}_m (t)$.

\subsection{Analytical results}

\subsubsection{Noiseless region ($K=0$)}

At $K= 0$, the probability of adopting the strategy of the role model described in Eq.\eqref{imitation} is a step function.
Thus, we can calculate
\begin{align}
    f(\Pi_{\mathrm{C}_m} - \Pi_{\mathrm{D}_{m'}})
    &= \theta (\Pi_{\mathrm{D}_{m'}} - \Pi_{\mathrm{C}_m})\nonumber\\
    &= \theta \left( r \frac{m'}{k+1} - r \frac{m+1}{k+1} + 1\right)\nonumber\\
    &= \theta \left( (m' - m - 1) + \frac{k+1}{r} \right).
\end{align}
Because $m$ and $m'$ are a non-negative integers, 
$f(\cdot)$ takes the same value in any $r$ satisfying $n-1 < (k+1) / r < n$ ($n \in \mathbb{N}$)
and changes discontinuously at $r=(k+1)/ n$.

\subsubsection{Noisy region ($K \gg 1$)}

If the noise $K$ is sufficiently large, we can approximate Eq.\eqref{imitation} as follows:
\begin{align}
    f(\Pi_x - \Pi_y)
    &\approx \frac{1}{2} - \frac{\Pi_x - \Pi_y}{4} K^{-1}.
\end{align}
Thus, by defining
\begin{align}
    W^{\mathrm{C}\to\mathrm{D}}_{m,\voter} &\equiv \frac{k-m}{k},\nonumber\\
    W^{\mathrm{D}\to\mathrm{C}}_{m,\voter} &\equiv \frac{m}{k},\nonumber\\
    W^{\mathrm{C}\to\mathrm{D}}_{m,\Pi} &\equiv \frac{k-m}{k} \left\langle \Pi_{\mathrm{C}_m} - \Pi_{\mathrm{D}} \right\rangle,\nonumber\\ 
    W^{\mathrm{D}\to\mathrm{C}}_{m,\Pi} &\equiv \frac{m}{k} \left\langle \Pi_{\mathrm{D}_m} - \Pi_{\mathrm{C}} \right\rangle,
\end{align}
we can divide the transition probability into the following two parts:
\begin{align}
    W^{s\to s'}_{m}
    &= \frac{1}{2} W^{s \to s'}_{m,\voter} 
    + \frac{1}{4K} W^{s \to s'}_{m,\Pi}.
\end{align}
Up to the zeroth order of $K^{-1}$, 
the dynamics of the model follows a transition rule such that 
each node changes its strategy (with $1/2$ probability) to that of a randomly selected neighbor.
This is essentially equivalent to the voter model\cite{liggett1985interacting,liggett1999stochastic}.
Using the AMEs and Eq.\eqref{CD_edge}, we can confirm that the fraction of cooperators remains constant:
\begin{align}
    \frac{\d}{\d t} \rho^{\mathrm{C}}
    &= \sum_m 
    \left[
        - \frac{1}{2} W^{\mathrm{C}\to\mathrm{D}}_{m,\voter} \rho^{\mathrm{C}}_m
        + \frac{1}{2} W^{\mathrm{D}\to\mathrm{C}}_{m,\voter} \rho^{\mathrm{D}}_m
    \right]\nonumber\\
    &= 
    \frac{1}{2k}
    \left[
        - \sum_m (k - m) \rho^{\mathrm{C}}_m
        + \sum_m m \rho^{\mathrm{D}}_m
    \right]
    = 0.
\end{align}
From this result, we can express the steady state of the voter model as $\rho^{*s}_{m}(\rho^{\mathrm{C}})$ as a function of $\rho^{\mathrm{C}}$.

Next, we consider the model dynamics up to the first order of $K^{-1}$.
If the noise is sufficiently large, the system first relaxes to the steady state of the voter model $\rho^{*s}_{m}(\rho^{\mathrm{C}})$.
Notice that the transition based on payoffs, $W^{s \to s'}_{m,\Pi}/4K$, is much smaller than the transition in the voter model, $W^{s \to s'}_{m,\voter}/2$.
Hence, the steady state of the voter model always holds: $\rho^{s}_{m} = \rho^{*s}_{m}(\rho^{\mathrm{C}})$.
Thus, the time derivative of $\rho^{\mathrm{C}}$ can be written as follows:
\begin{align}
    \frac{\d}{\d t} \rho^{\mathrm{C}}
    =& 
    \sum_m 
    \left[
        -W^{\mathrm{C}\to\mathrm{D}}_m \rho^{*\mathrm{C}}_m
        +W^{\mathrm{D}\to\mathrm{C}}_m \rho^{*\mathrm{D}}_m
    \right]\nonumber\\
    =& 
    \frac{1}{4K} \sum_m 
    \left[
        -W^{\mathrm{C}\to\mathrm{D}}_{m,\Pi} \rho^{*\mathrm{C}}_m
        +W^{\mathrm{D}\to\mathrm{C}}_{m,\Pi} \rho^{*\mathrm{D}}_m
    \right]\nonumber\\
    =& 
    \frac{1}{2K k \sum_{m''} m'' \rho^{*\mathrm{D}}_{m''}}\nonumber\\
    &\times 
    \sum_m \sum_{m'} m (k - m')\rho^{*\mathrm{D}}_{m} \rho^{*\mathrm{C}}_{m'}
    \left( \Pi_{\mathrm{C}_{m'}} - \Pi_{\mathrm{D}_m} \right)\nonumber\\
    =&
    \frac{1}{2Kk(k+1)}
    \left[
        r \sum_m \left[
            m(k - m) \rho^{*\mathrm{C}}_m
            - m^2 \rho^{*\mathrm{D}}_m
        \right]
    \right.\nonumber\\
    &
    \left.
        + [r - (k+1)] \sum_m (k - m) \rho^{*\mathrm{C}}_m
    \right].\nonumber
\end{align}
Since $\rho^{*s}_{m}(\rho^{\mathrm{C}})$ satisfy (see appendix \ref{voter model})
\begin{align}
    \sum_m m (k - m) \rho^{*\mathrm{C}}_m &= \sum_m m^2 \rho^{*\mathrm{D}}_m,
    \label{voter_relationship}
\end{align}
we obtain
\begin{align}
    \frac{\d}{\d t} \rho^{\mathrm{C}}
    &=
    \frac{1}{2Kk(k+1)}
    [r - (k+1)] \sum_m (k - m) \rho^{*\mathrm{C}}_m (\rho^{\mathrm{C}}).
\end{align}
Therefore, if the noise $K$ is sufficiently large, the (C+D) phase does not exist, and the boundary between the C and D phases is $r=k+1$.

\subsubsection{Condition for the survival of cooperators}

We roughly estimate the condition for which cooperators can survive.
The following equation is obtained from the AMEs and Eq.\eqref{CD_edge}:
\begin{align}
    \frac{\d}{\d t} \rho^{\mathrm{C}}
    =& \frac{1}{k \sum_{m''} m'' \rho^{\mathrm{D}}_{m''}}
    \sum_{m=1}^{k} \sum_{m'=0}^{k-1} m (k - m')
    \rho^{\mathrm{C}}_{m'} \rho^{\mathrm{D}}_m\nonumber\\
    &\times 
    \tanh \left(\frac{\Pi_{\mathrm{C}_{m'}} - \Pi_{\mathrm{D}_m}}{2K} \right).
\end{align}
Since $\Pi_{\mathrm{C}_{k-1}} - \Pi_{\mathrm{D}_1} = r(k-1)/(k+1) - 1$, for $r < (k+1)/(k-1)$ 
all the terms on the right side of the above equation become negative, and the cooperators become extinct.
That is, the lower boundary for the C and (C+D) phases must satisfy $r > (k+1)/(k-1)$.

\subsubsection{Condition for the survival of defectors and power relaxation}

In the parameter region where defectors manage to survive, we expect that clusters of defectors exist separated from each other in the sea of C.
We first seek the conditions under which an isolated D-cluster can survive, ignoring interactions among D-clusters.
Since defectors do not contribute to the public good, a D-node has the largest payoff when it is completely surrounded by C-nodes.
Thus, no D-cluster of any size can survive under the condition where even an isolated D-node cannot survive.

Now, we consider the region where $r > k+1$.
In this region, the payoff of an isolated defector is larger than that of its neighboring cooperator.
In contrast, the payoff of a defector in a cluster consisting of two or more defectors is always smaller than that of its neighboring cooperator.
If there is no noise (i.e., $K=0$), a node always imitates the strategy of a role model if and only if the payoff of the role model is larger than its own.
Therefore, D-clusters oscillate in size between 1 and 2.
D-clusters also oscillate if the noise is not negligible, though isolated D-nodes disappear with some probability.

Next, we take into account the interactions among D-clusters.
We consider the region $r > k+1$ and $K=0$.
In the following discussion, we refer to a D-cluster of size 1 or 2 as a tiny D-cluster.
Since the node that updates its strategy is randomly chosen, the D-node in a D-cluster of size 2 that switch to C is also randomly determined.
This implies that each tiny D-cluster moves randomly on the network like an inchworm.
When a tiny D-cluster comes next to another one, these clusters merge to form one cluster, which then shrinks to form a tiny D-cluster, with a probability larger than zero.
Therefore, the dynamics can be regarded as a coalescing random walk.
In this case, the problem of finding the relaxation time is equivalent to finding the coalescence time for the corresponding coalescing random walk.
On a finite regular random graph, the coalescence continues until there exists only one D-cluster.
Previous investigations \cite{cooper2010multiple,cooper2013coalescing} have revealed that the coalescence time is $\O (N)$.
On an infinite regular random graph (or the Bethe lattice), any two random walkers do not meet with a probability larger than zero, and the coalescence time diverges.
Therefore, in the AMEs case also, the region $r > k+1$ and $K=0$ belongs to the (C+D) phase, but the steady state is different from that in the finite case.

We can roughly estimate the relaxation time on a finite regular random graph for $r > k+1$ and $K=0$ by considering the evolution of the number of tiny D-clusters (or random walkers), represented by $x(t)$.
The probability that either tiny D-clusters move in one strategy updating is proportional to the number of tiny D-clusters $x(t)$.
Furthermore, we estimate the probability that a tiny D-cluster visits a certain node in one step as $1/N$.
When a tiny D-cluster visits a node near another tiny D-cluster, these clusters coalesce into a tiny D-cluster with a certain probability.
That is, the probability that a random walker disappears in one step is proportional to $(x(t)-1)/N$.
Therefore, the evolution of $x(t)$ can be approximated as follows:
\begin{align}
    \frac{\d}{\d t} x(t) 
    &\approx - \alpha x \frac{x-1}{N}
    = \frac{\alpha}{N} x (1-x),
\end{align}
where $\alpha$ is a constant.
This is a logistic equation and can be solved:
\begin{align}
    x(t) 
    &= \frac{x(0)}{x(0) + (1 - x(0)) \e^{- \alpha t / N}}
    \approx \frac{1}{1 - \e^{- \alpha t / N}}.
\end{align}
Note that we assume $x(0) \gg 1$.
Since $x(t) \to 1$ in the limit $t \to \infty$, eventually only one D-cluster will survive.
Additionally, the number of D-clusters at time $\tau \equiv t/N$ is independent of the number of nodes $N$.
This result is consistent with the fact that the coalescence time is $\mathcal{O}(N)$.
If $\tau$ is small enough, we can estimate $x(t) \approx \alpha^{-1} \tau^{-1}$ and $\rho^{\mathrm{D}}(t) \propto x(t)/N = \alpha^{-1} t^{-1}$.
That is, the fraction of defectors is independent of $N$ and decreases as $t^{-1}$.

\subsection{Numerical calculations}

\begin{figure}[hbtp]
    \includegraphics[width=1.0\linewidth]{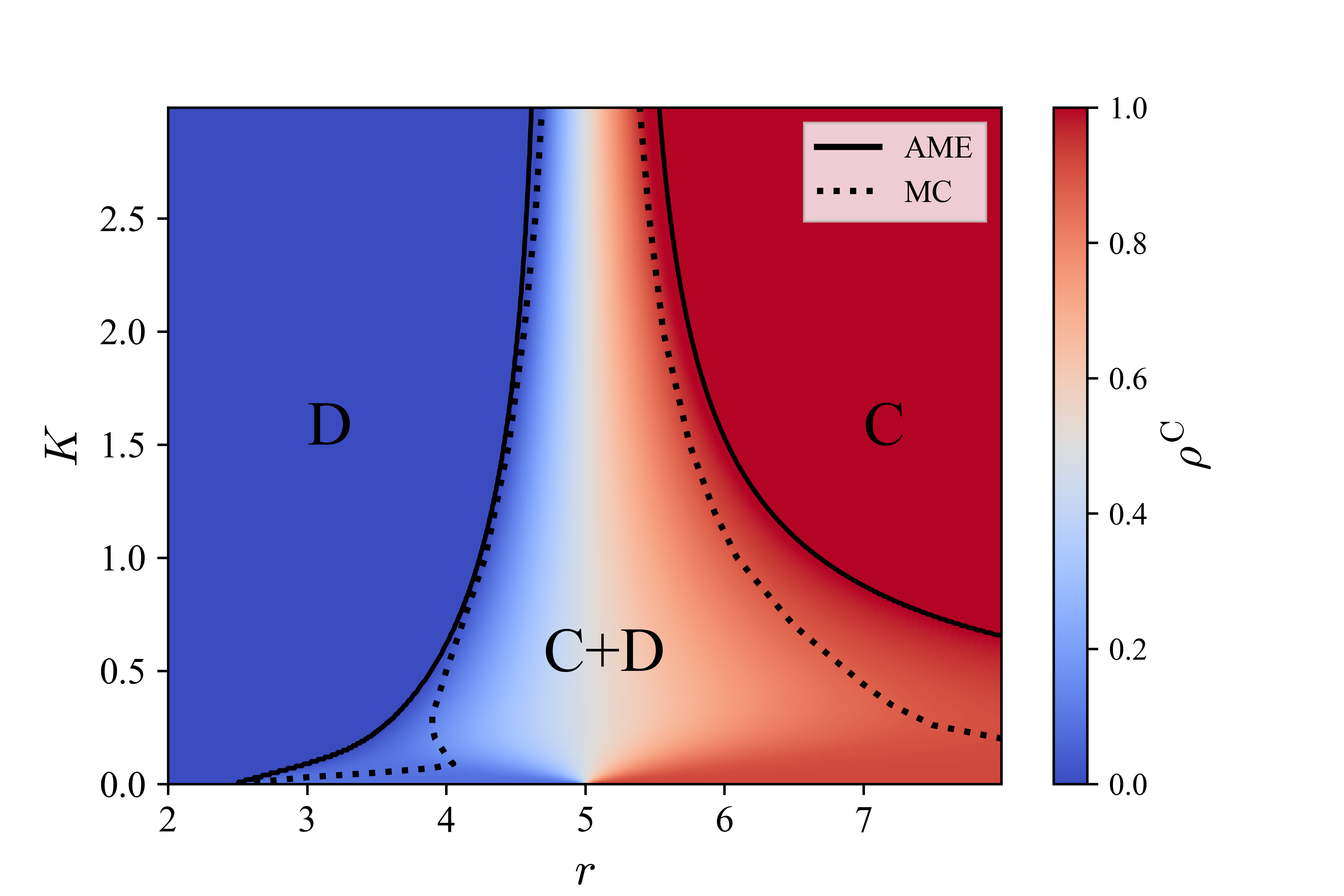}
    \caption{
        Phase diagram with degree $k=4$.
        The solid lines and the color map are the results obtained by numerical calculations of the approximate master equations 
        (AMEs)  using the Euler method,
        with the time step $1/10$, computed up to $t=5\times 10^3$.
        We define $\rho^{\mathrm{C}} < 10^{-5} $ as the C phase (and the same for the D phase).
        The dotted lines indicate the phase boundary obtained from the Monte Carlo (MC) simulations.
        The number of nodes $N=10^5$, $5\times 10^3$MCS.
        \label{PhaseDiagram}
    }
\end{figure}

FIG.\ref{PhaseDiagram} shows the phase diagrams obtained by using the AMEs and the MC simulation.
We ran the MC simulations on a network with $N=10^5$ and $k=4$ up to $5 \times 10^3$ Monte Carlo steps (MCS; $N$ strategy updating per MCS).
For both phase diagrams, in the absence of noise (i.e., $K=0$), the boundary between the D and (C+D) phases is $r = (k+1)/2 = 2.5$, 
and defectors survive even in the limit $r \to \infty$.
As the noise increases, the phase boundary approaches $r= k+1 = 5$.
These results are consistent with those obtained analytically.
While cooperators go extinct in a well-mixed population (or in the mean-field approximation), 
they can survive in some regions when network structure is considered (network reciprocity\cite{nowak1992evolutionary,nowak2006five}).
This is because structured populations allow heterogeneous distributions, enabling cooperators to cluster.
\begin{figure}[hbtp]
    \includegraphics[width=1.0\linewidth]{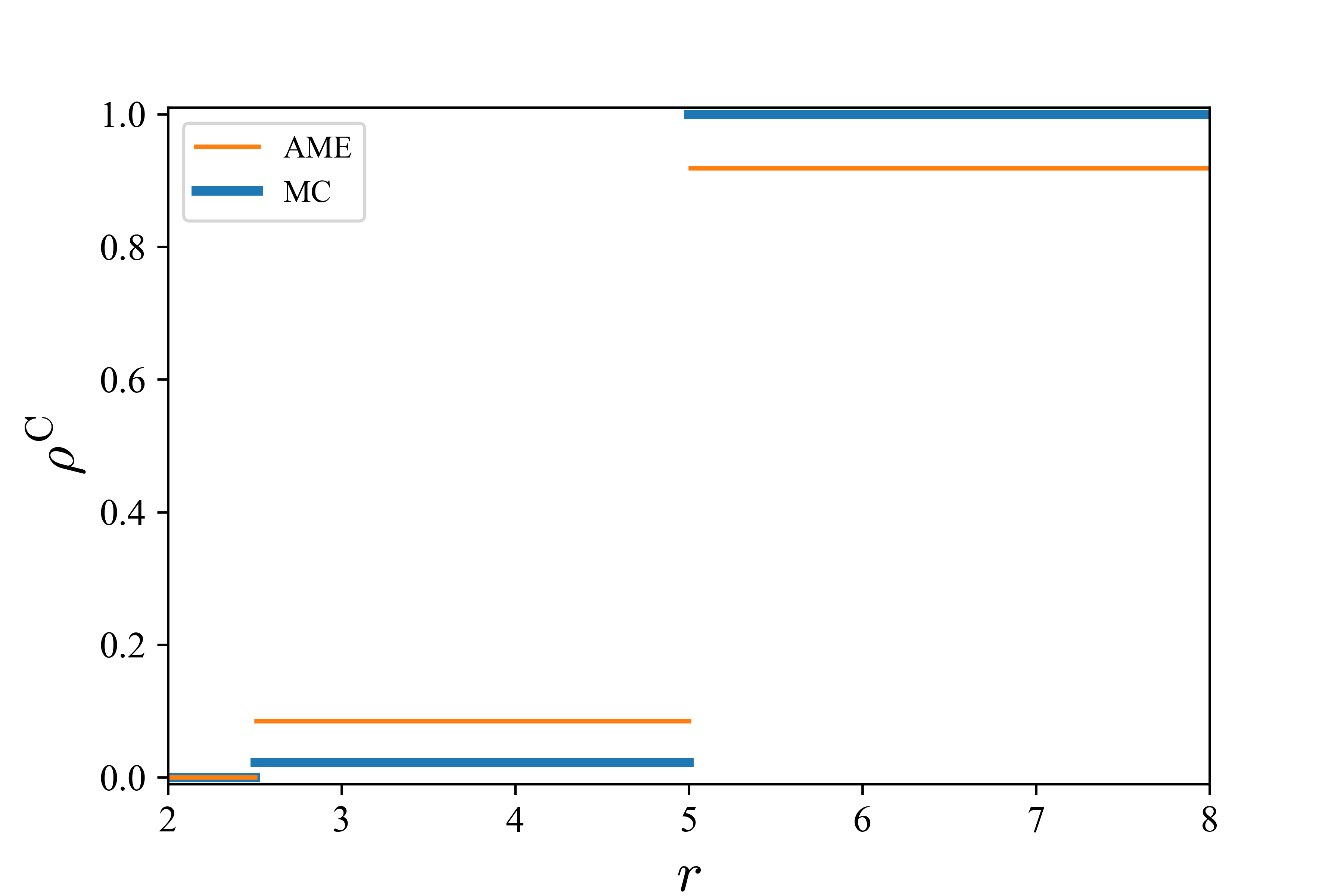}
    \includegraphics[width=1.0\linewidth]{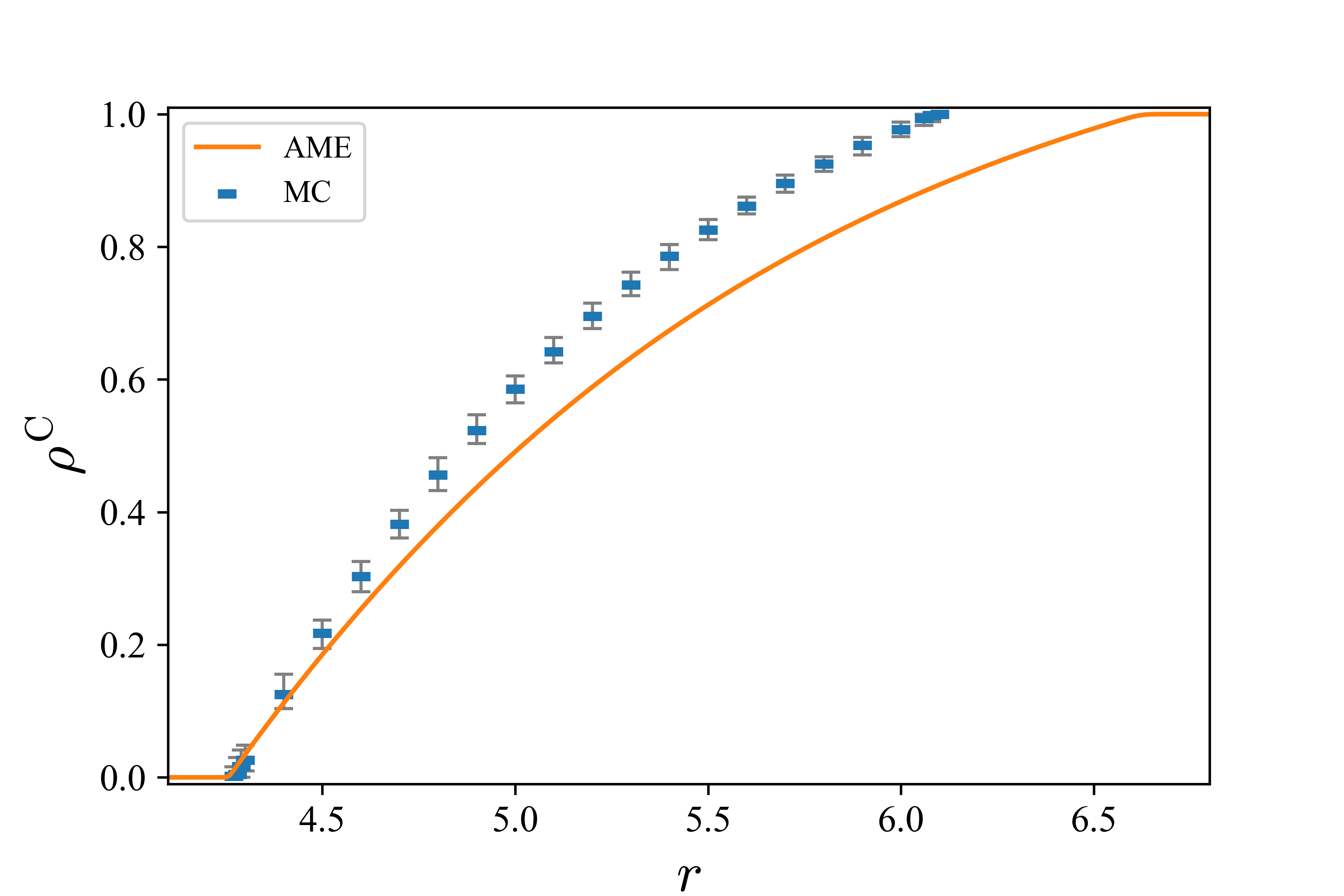}
    \caption{
        Fraction of cooperators at $K=0$ (upper) and $K=1$ (lower).
        The orange lines and the blue lines (or marks) denote the results obtained by the AMEs and the MC simulations, respectively.
        The MCS were taken up to $5\times 10^3$ ($1.5\times 10^5$ in $K=0$ and $r > 5$), and the last $10^3$ MCS results were averaged.
        In addition, independent simulations were run $10$ times and averaged.
        The error bars represent the maximum and minimum values among all samples.
        At $K=0$, the variance is so small that the error bar overlaps the line of the average, and hence, this case is not shown.
        \label{K_rhoC}
    }
\end{figure}

\begin{figure}[hbtp]
    \includegraphics[width=1.0\linewidth]{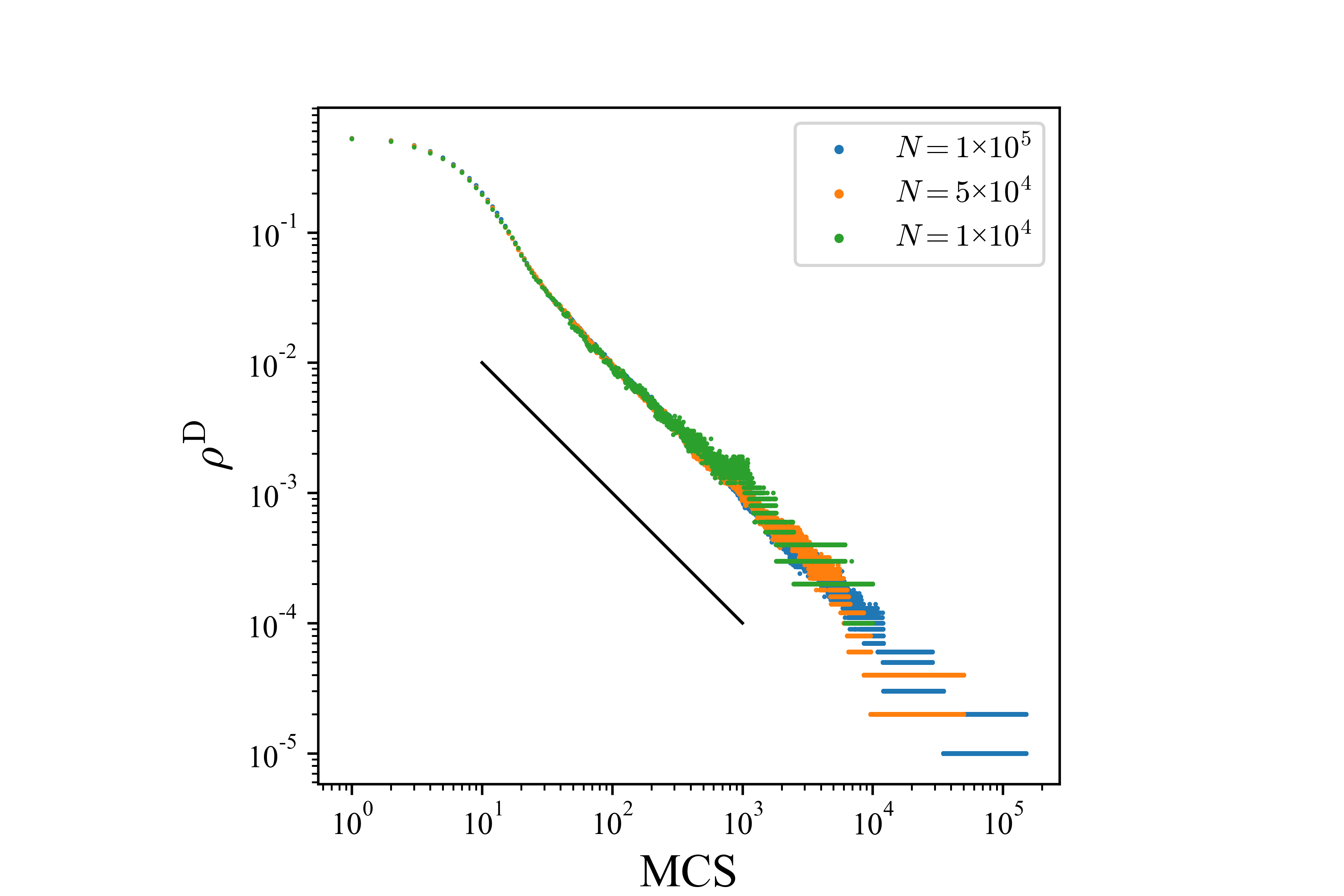}
    \caption{
        Time evolution of $\rho^{\mathrm{D}}$ in $r > 5$ and $K=0$, and the number of nodes $N=10^5$ (blue), $5 \times 10^4$ (orange), $10^4$ (green).
        The solid line has a slope of $-1$.
        \label{CoalescingTime}
    }
\end{figure}

The $r$ dependance of $\rho^{\mathrm{C}}$ at $K=0$ and $K=1$ is shown in FIG.\ref{K_rhoC}.
At $K=0$, the fraction of cooperators changes discontinuously at $r=5,\ 2.5$ but stays constant in the other $r$.
FIG.\ref{CoalescingTime} shows the time evolution of $\rho^{\mathrm{C}}$ in $r > k+1$ and $K=0$.
As revealed from the analytical results, $\rho^{\mathrm{C}} $ decreases as $t^{-1}$, independent of $N$, until only one tiny D-cluster remains.
Conversely, no power relaxation was seen in the AMEs.

\section{Summary and Discussion}

We derived the AMEs for the spatial public goods game and found that the results obtained by the AMEs are mostly qualitatively consistent with those by the MC simulations.
Furthermore, we confirmed that discontinuous phase transitions occur for $K=0$ and $K\to\infty$ and that the transition points are quantitatively consistent with the AMEs and MC simulation results.
These transition points are also consistent with analytically obtained results.
In the MC simulations, the phase boundary is sharp around low $K$ and $r \approx 4$, but such a sharp boundary is not seen in the AMEs results
because of the finite size effects and insufficient approximation accuracy of the AMEs.

At $K \gg 1$, we clarified that the steady state of the voter model drastically affects the phase boundary.
We note that the steady state of the voter model depends on whether the network is finite or infinite, or on the dimension of the network.
For a finite network or an infinite network with one or two dimensions,
the initial state (rather than the synergy factor $r$) makes a larger contribution to determining the absorbing state (i.e., $\rho^{\mathrm{C}}=0$ or $1$) that the system will eventually reach.
Conversely, in the AMEs that corresponds to the Bethe lattice, the system will always reach the absorbing state $\rho^{\mathrm{C}}=1$ when $r>k+1$, 
even if there are more defectors than cooperators in the initial state.

We showed that the value of $\rho^{\mathrm{C}}$ varies discontinuously in the noiseless region.
The discontinuity is due to the shape of the function representing the probability of strategy updating and the discrete number of cooperators;
therefore, networks with any degree distributions or models with nonuniform group sizes should also show discontinuities,
as long as they follow the same probability $f$.

As described above, our approach is useful for analyzing the model in detail.
Additionally, our method can be applied to other models (see appendix \ref{application}).
It can be easily extended to models that follow another public goods function $F(N_{\mathrm{C}})$
or to models with multiple strategy update rules (e.g., Javarone et al.\cite{javarone2016conformity}).
The method can also be applied to complex many-body interaction models that depend on strategies of the next-nearest nodes or more distant nodes.
Furthermore, the AMEs have been formulated for arbitrary order distributions\cite{gleeson2013binary}.
Thus, this method can be applied to models on heterogeneous networks where extended mean-field approximations considering blocks of nodes\cite{szabo2005phase,vukov2006cooperation} are not available.

\section*{Acknowledgement}
We would like to thank Takehisa Hasegawa for useful discussions.

\appendix

\onecolumngrid

\section{Derivation of the AMEs}
\label{deviation of AME}

\begin{figure}[hbtp]
    \begin{minipage}[b]{0.40\linewidth}
        \centering
    \includegraphics[width=1.0\linewidth]{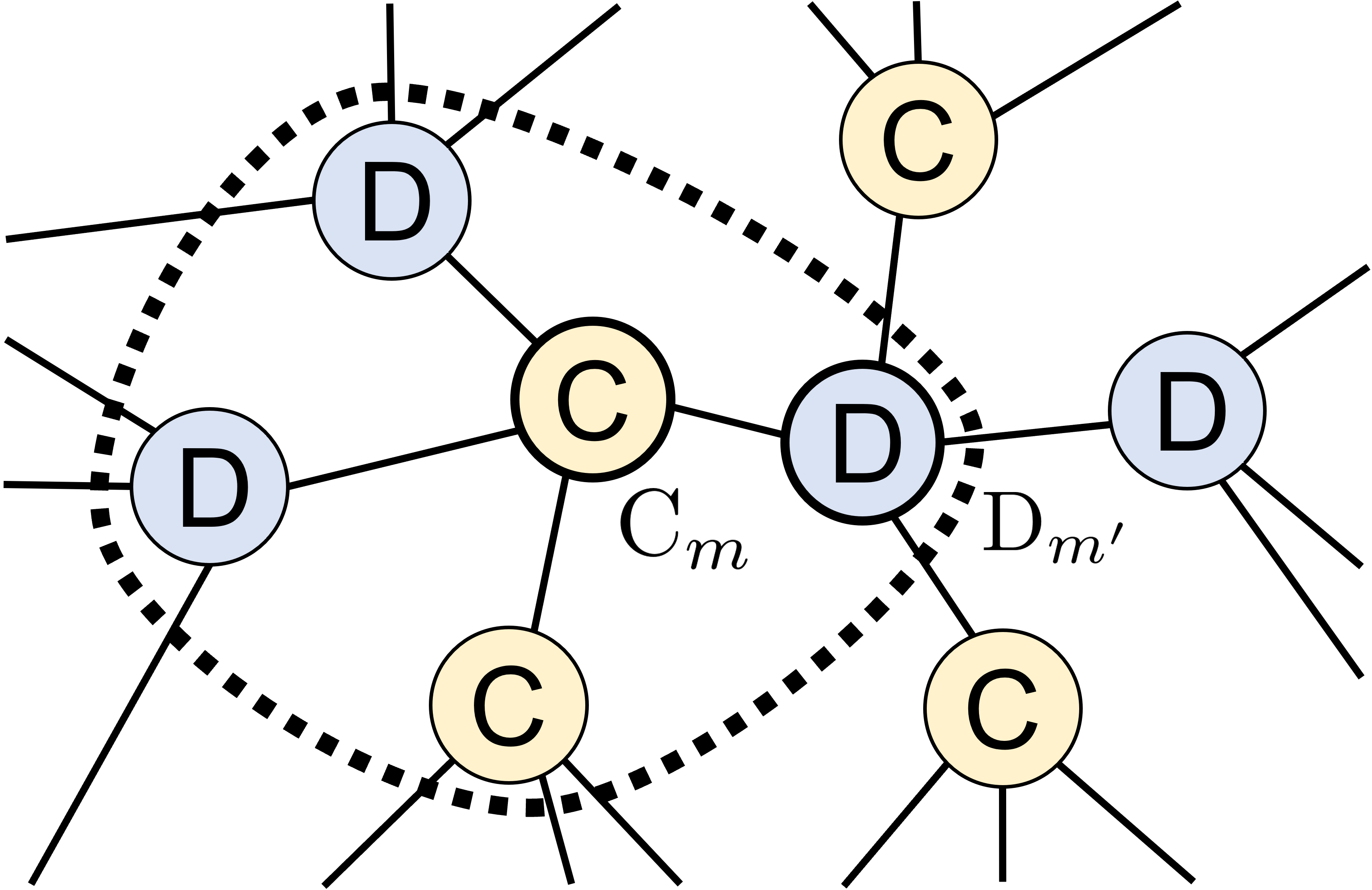}
    \caption{
        Schematic showing the $\mathrm{C}_m$ node imitating the strategy of the $\mathrm{D}_{m'}$ node. 
        The dotted enclosure represents the group containing the $\mathrm{C}_m$ node.
        \label{AMEgroup}
    }
    \end{minipage}
    \begin{minipage}[b]{0.45\linewidth}
        \centering
        \includegraphics[width=1.0\linewidth]{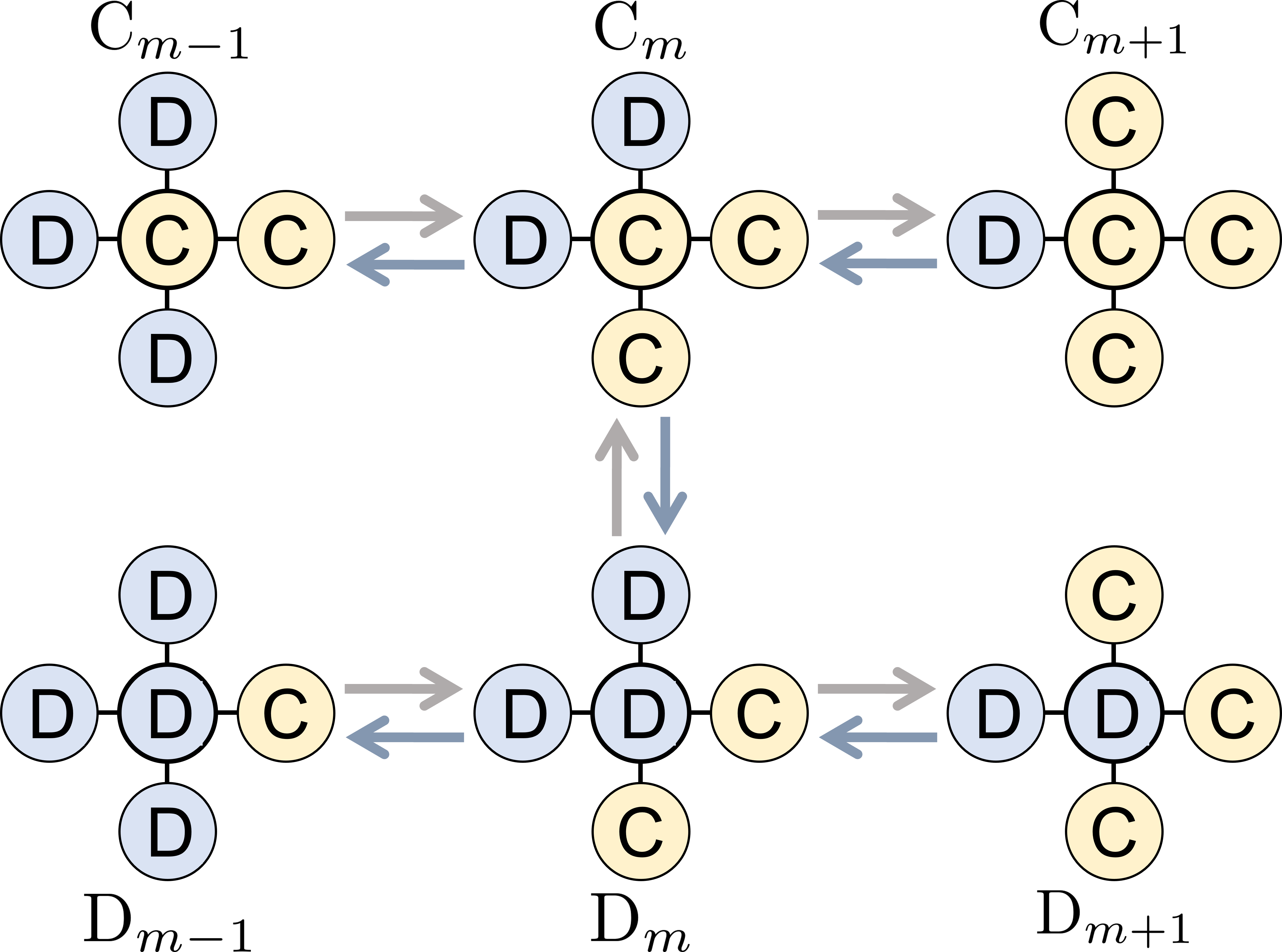}
        \caption{
        Schematic of transitions in the AME.
        \label{AMEtransition}
    }
    \end{minipage}
\end{figure}

The time evolution of the fraction of $s_m$ nodes is expressed as follows (see FIG.\ref{AMEtransition}):
\begin{align}
    \frac{\d}{\d t} \rho^{\mathrm{C}}_m (t)
    =&
    - W(\mathrm{C}_m \to \mathrm{D}_m) \rho^{\mathrm{C}}_m
    + W(\mathrm{D}_m \to \mathrm{C}_m) \rho^{\mathrm{D}}_m\nonumber\\
    &- W(\mathrm{C}_m \to \mathrm{C}_{m+1}) \rho^{\mathrm{C}}_m
    + W(\mathrm{C}_{m-1} \to \mathrm{C}_m) \rho^{\mathrm{C}}_{m-1}
    - W(\mathrm{C}_m \to \mathrm{C}_{m-1}) \rho^{\mathrm{C}}_m
    + W(\mathrm{C}_{m+1} \to \mathrm{C}_m) \rho^{\mathrm{C}}_{m+1},\\
    \frac{\d}{\d t} \rho^{\mathrm{D}}_m (t)
    =&
    - W(\mathrm{D}_m \to \mathrm{C}_m) \rho^{\mathrm{D}}_m
    + W(\mathrm{C}_m \to \mathrm{D}_m) \rho^{\mathrm{C}}_m\nonumber\\
    &- W(\mathrm{D}_m \to \mathrm{D}_{m+1}) \rho^{\mathrm{D}}_m
    + W(\mathrm{D}_{m-1} \to \mathrm{D}_m) \rho^{\mathrm{D}}_{m-1}
    - W(\mathrm{D}_m \to \mathrm{D}_{m-1}) \rho^{\mathrm{D}}_m
    + W(\mathrm{D}_{m+1} \to \mathrm{D}_m) \rho^{\mathrm{D}}_{m+1},
\end{align}
where $W(s_m \to s'_{m'})$ is the transition probability from $s_m$ to $s'_{m'}$.
We first derive the expression of $W(\mathrm{C}_m \to \mathrm{D}_m)$ (see also FIG.\ref{AMEgroup}).
When a $\mathrm{C}_m$ node is chosen as the node that updates its strategy, a defector is chosen as the role model with probability $(k-m)/m$ 
and the $\mathrm{C}_m$ node imitates the strategy of the role model with probability $f$:
\begin{align}
    W^{\mathrm{C}\to\mathrm{D}}_{m}
    &\equiv W(\mathrm{C}_m \to \mathrm{D}_m)
    = \frac{k-m}{k} \left\langle f(\Pi_{\mathrm{C}_m} - \Pi_{\mathrm{D}}) \right\rangle
    = \frac{k-m}{k} \sum_{m'=0}^k p_{\mathrm{C}}^{\mathrm{D}}(m') f(\Pi_{\mathrm{C}_m} - \Pi_{\mathrm{D}_{m'}}),
    \label{W_CtoD}
\end{align}
where we give the probability $f$ by the expected value, and denote $p_{s}^{s'}(m')$ as the probability that a node is $s'_{m'}$ under the condition that it adopts strategy $s'$ and is adjacent to $s$ node:
\begin{align}
    p_{\mathrm{C}}^{\mathrm{C}}(m')
    = \frac{m' \rho^{\mathrm{C}}_{m'}}{\sum_{m} m \rho^{\mathrm{C}}_{m}},&&
    p_{\mathrm{C}}^{\mathrm{D}}(m')
    = \frac{m' \rho^{\mathrm{D}}_{m'}}{\sum_{m} m \rho^{\mathrm{D}}_{m}},&&
    p_{\mathrm{D}}^{\mathrm{C}}(m')
    = \frac{(k - m') \rho^{\mathrm{C}}_{m'}}{\sum_{m} (k - m) \rho^{\mathrm{C}}_{m}},&&
    p_{\mathrm{D}}^{\mathrm{D}}(m')
    = \frac{(k - m') \rho^{\mathrm{D}}_{m'}}{\sum_{m} (k - m) \rho^{\mathrm{D}}_{m}}.
\end{align}
We approximated $p_{s}^{s'}(m'|m)$ by $p_{s}^{s'}(m')$ in Eq.\eqref{W_CtoD}, where $p_{s}^{s'}(m'|m)$ is the probability that a node is $s'_{m'}$ under the condition that it adopts strategy $s'$ and is adjacent to $s_{m}$ node.
If $m$ and $m'$ are correlated, they are not equal. 
In networks with a large average clustering coefficient, there is a positive correlation between $m$ and $m'$ because there are nodes adjacent to both nodes.
Therefore, this approximation is more accurate for networks with fewer loops.
Similarly, we derive
\begin{align}
    W^{\mathrm{D}\to\mathrm{C}}_{m}
    &\equiv W(\mathrm{D}_m \to \mathrm{C}_m)
    = \frac{m}{k} \left\langle f(\Pi_{\mathrm{D}_m} - \Pi_{\mathrm{C}}) \right\rangle
    = \frac{m}{k} \sum_{m'=0}^k p_{\mathrm{D}}^{\mathrm{C}}(m') f(\Pi_{\mathrm{D}_m} - \Pi_{\mathrm{C}_{m'}}).
\end{align}
We define $\beta^{\mathrm{C}}$ ($\gamma^{\mathrm{C}}$) as the probability that a defector (cooperator) adjacent to a cooperator changes the strategy to C (D):
\begin{align}
    \beta^{\mathrm{C}} = \sum_{m'} p_{\mathrm{C}}^{\mathrm{D}}(m') \ W^{\mathrm{D} \to \mathrm{C}}_{m'},&&
    \gamma^{\mathrm{C}} = \sum_{m'} p_{\mathrm{C}}^{\mathrm{C}}(m') \ W^{\mathrm{C} \to \mathrm{D}}_{m'}.
\end{align}
We can express the transition probability as follows:
\begin{align}
    W(\mathrm{C}_m \to \mathrm{C}_{m+1}) &= \beta^{\mathrm{C}} (k - m),&
    W(\mathrm{C}_{m-1} \to \mathrm{C}_m) &= \beta^{\mathrm{C}} (k - m + 1),\nonumber\\
    W(\mathrm{C}_m \to \mathrm{C}_{m-1}) &= \gamma^{\mathrm{C}} m,&
    W(\mathrm{C}_{m+1} \to \mathrm{C}_m) &= \gamma^{\mathrm{C}} (m + 1).
\end{align} 
We define $\beta^{\mathrm{D}}$ and $\gamma^{\mathrm{D}}$ similarly.
By using these terms, we can derive the AMEs for the public goods game, Eq.\eqref{AME}.

\section{Relation satisfied in the steady state of the voter model}
\label{voter model}

To obtain the AMEs for the voter model, we only replace $W^{s\to s'}_{m}$ with $W^{s\to s'}_{m,\voter}$.
We calculate
\begin{align}
    \frac{\d}{\d t}& \sum_m m \rho^{\mathrm{C}}_m\nonumber\\
    =&
    - \frac{1}{k} \sum_m m (k-m) \rho^{\mathrm{C}}_m
    + \frac{1}{k} \sum_m m^2 \rho^{\mathrm{D}}_m \nonumber\\
    &
    - \frac{1}{k} \frac{\sum_{m'} m'^2 \rho^{\mathrm{D}}_{m'}}{\sum_{m'} m' \rho^{\mathrm{D}}_{m'}}
    \sum_m m (k - m) \rho^{\mathrm{C}}_m
    + \frac{1}{k} \frac{\sum_{m'} m'^2 \rho^{\mathrm{D}}_{m'}}{\sum_{m'} m' \rho^{\mathrm{D}}_{m'}} 
    \sum_m m (k - m + 1) \rho^{\mathrm{C}}_{m-1} \nonumber\\
    &
    - \frac{1}{k} \frac{\sum_{m'} m' (k-m') \rho^{\mathrm{C}}_{m'}}{\sum_{m'} m' \rho^{\mathrm{C}}_{m'}} 
    \sum_m m^2 \rho^{\mathrm{C}}_m
    + \frac{1}{k} \frac{\sum_{m'} m' (k-m') \rho^{\mathrm{C}}_{m'}}{\sum_{m'} m' \rho^{\mathrm{C}}_{m'}} 
    \sum_m m (m+1) \rho^{\mathrm{C}}_{m+1}\nonumber\\
    =&
    - \frac{1}{k} \sum_m m (k-m) \rho^{\mathrm{C}}_m
    + \frac{1}{k} \sum_m m^2 \rho^{\mathrm{D}}_m \nonumber\\
    &
    - \frac{1}{k} \frac{\sum_{m'} m'^2 \rho^{\mathrm{D}}_{m'}}{\sum_{m'} m' \rho^{\mathrm{D}}_{m'}}
    \sum_{m} m (k - m) \rho^{\mathrm{C}}_m
    + \frac{1}{k} \frac{\sum_{m'} m'^2 \rho^{\mathrm{D}}_{m'}}{\sum_{m'} m' \rho^{\mathrm{D}}_{m'}} 
    \sum_{m''=0}^{k-1} (m''+1) (k - m'') \rho^{\mathrm{C}}_{m''} \nonumber\\
    &
    - \frac{1}{k} \frac{\sum_{m'} m' (k-m') \rho^{\mathrm{C}}_{m'}}{\sum_{m'} m' \rho^{\mathrm{C}}_{m'}} 
    \sum_{m} m^2 \rho^{\mathrm{C}}_m
    + \frac{1}{k} \frac{\sum_{m'} m' (k-m') \rho^{\mathrm{C}}_{m'}}{\sum_{m'} m' \rho^{\mathrm{C}}_{m'}} 
    \sum_{m''=1}^{k} (m''-1) m'' \rho^{\mathrm{C}}_{m''}\nonumber\\
    =&
    - \frac{1}{k} \sum_m m (k-m) \rho^{\mathrm{C}}_m
    + \frac{1}{k} \sum_m m^2 \rho^{\mathrm{D}}_m
    + \frac{1}{k} \frac{\sum_{m'} m'^2 \rho^{\mathrm{D}}_{m'}}{\sum_{m'} m' \rho^{\mathrm{D}}_{m'}} 
    \sum_{m''} (k - m'') \rho^{\mathrm{C}}_{m''}
    - \frac{1}{k} \frac{\sum_{m'} m' (k-m') \rho^{\mathrm{C}}_{m'}}{\sum_{m'} m' \rho^{\mathrm{C}}_{m'}} 
    \sum_{m''} m'' \rho^{\mathrm{C}}_{m''}\nonumber\\
    =&
    \frac{2}{k} 
    \left[
        \sum_m m^2 \rho^{\mathrm{D}}_m - \sum_m m (k-m) \rho^{\mathrm{C}}_m
    \right],
    \label{voter_m_C}
\end{align}
where we use Eq.\eqref{CD_edge}.
Since the left side is zero in the steady state, we obtain 
\begin{align}
    \sum_m m (k - m) \rho^{\mathrm{C}}_m &= \sum_m m^2 \rho^{\mathrm{D}}_m.
\end{align}
Likewise, from 
\begin{align}
    \frac{\d}{\d t} \sum_m m^2 \rho^{\mathrm{C}}_m +  \frac{\d}{\d t} \sum_m m^2 \rho^{\mathrm{D}}_m = 0
\end{align}
in the steady state, we can obtain
\begin{align}
    0 
    =& \sum_m m^2 \rho^{\mathrm{D}}_m 
    \left[ 
        \frac{\sum_m m^2 \rho^{\mathrm{D}}_m}{\sum_m m \rho^{\mathrm{D}}_m} 
        - \frac{\sum_m m^2 \rho^{\mathrm{C}}_m}{\sum_m m \rho^{\mathrm{C}}_m} 
        + 1
    \right]
    + \sum_m m (k-m) \rho^{\mathrm{D}}_m
    \left[ 
        \frac{\sum_m m (k-m) \rho^{\mathrm{D}}_m}{\sum_m (k-m) \rho^{\mathrm{D}}_m} 
        - \frac{\sum_m m (k-m) \rho^{\mathrm{C}}_m}{\sum_m (k-m) \rho^{\mathrm{C}}_m} 
        + 1
    \right].
    \label{voter_relationship2}
\end{align}

\section{Applications to other models}
\label{application}

Previous studies\cite{szolnoki2009topology,javarone2016role} have examined the model in which each node belongs not only to a group in which the center node is itself 
but also to $k$ groups in which the center node is its neighbor.
The overall payoff of each node is the sum of the payoffs in each group.
In this case, we replace the expected value of $f$ with
\begin{align}
    \left\langle f(\Pi_{\mathrm{C}_m} - \Pi_{\mathrm{D}}) \right\rangle
    =& 
    \sum_{m' =0}^{k} 
    \sum_{\{m_{x}\}=0}^{k}
    \sum_{\{m_{y}\}=0}^{k}
    p_{\mathrm{C}}^{\mathrm{D}} (m')
    \left(
        \prod_{i=1}^m p_{\mathrm{C}}^{\mathrm{C}}(m_{xi})
    \right)
    \left(
        \prod_{j=m+1}^{k-1} p_{\mathrm{C}}^{\mathrm{D}}(m_{xj})
    \right)
    \left(
        \prod_{s=1}^{m'-1} p_{\mathrm{D}}^{\mathrm{C}}(m_{ys})
    \right)
    \left(
        \prod_{t=m'}^{k-1} p_{\mathrm{D}}^{\mathrm{D}}(m_{yt})
    \right)\nonumber\\
    &\times
    f \left[
        \Pi_{\mathrm{C}}(2m + m' + m_{x1} + \cdots + m_{x(k-1)} + 1)
        - \Pi_{\mathrm{D}}(2m' + m + m_{y1} + \cdots + m_{y(k-1)}) 
    \right],\\
    \left\langle f(\Pi_{\mathrm{D}_m} - \Pi_{\mathrm{C}}) \right\rangle
    =& 
    \sum_{m' =0}^{k}
    \sum_{\{m_{x}\}=0}^{k}
    \sum_{\{m_{y}\}=0}^{k}
    p_{\mathrm{D}}^{\mathrm{C}}(m')
    \left(
        \prod_{i=1}^{m-1} p_{\mathrm{D}}^{\mathrm{C}}(m_{xi})
    \right)
    \left(
        \prod_{j=m}^{k-1} p_{\mathrm{D}}^{\mathrm{D}}(m_{xj})
    \right)
    \left(
        \prod_{s=1}^{m'} p_{\mathrm{C}}^{\mathrm{C}}(m_{ys})
    \right)
    \left(
        \prod_{t=m'+1}^{k-1} p_{\mathrm{C}}^{\mathrm{D}}(m_{yt})
    \right)\nonumber\\
    &\times 
    f \left[
        \Pi_{\mathrm{D}}(2m + m' + m_{x1} + \cdots + m_{x(k-1)})
        - \Pi_{\mathrm{C}}(2m' + m + m_{y1} + \cdots + m_{y(k-1)} + 1) 
    \right],
\end{align}
where
\begin{align}
    \Pi_{\mathrm{C}}(N_{\mathrm{C}}^{\text{all}}) 
    \equiv r \frac{N_{\mathrm{C}}^{\text{all}}}{k+1} - (k + 1),&&
    \Pi_{\mathrm{D}}(N_{\mathrm{C}}^{\text{all}}) 
    \equiv r \frac{N_{\mathrm{C}}^{\text{all}}}{k+1}.
\end{align}
In this model, the points of discontinuity at $K=0$ are $r=(k+1)^2/n$ ($n \in \mathbb{N}$).
At $K \gg 1$, we can obtain
\begin{align}
    \frac{\d}{\d t} \rho^{\mathrm{C}}
    &=
    \frac{k+1}{2Kk(k-1)}
    [r - (k+1)] \sum_m (k - m) \rho^{*\mathrm{C}}_m
\end{align}
by using \eqref{voter_relationship} and \eqref{voter_relationship2}.
Thus, as the noise increases, the phase boundary approaches $r= k+1$.
Further, the lower boundary for the C and (C+D) phases must satisfy $r > (k+1)/(k-1)$.
Even an isolated D-node cannot survive in $(k+1)^2/3 < r$, and the coalescing random walk occurs in $(k+1)^2/4 < r < (k+1)^2 /3$ and $K=0$.
We confirmed that the above analytical results are correct by computing the AMEs numerically and the MC simulations.
Furthermore, these analytical results are mostly consistent with the results of previous studies that employed square lattices as networks.
However, Szolnoki et.al.\cite{szolnoki2009topology} identified the powerfully relaxing region (i.e., $(k+1)^2/4 < r < (k+1)^2 /3$ and $K=0$) as the C phase instead of the (C+D) phase.

The public goods game models group interactions, while the prisoner's dilemma game models pair interactions.
In the latter model, a cooperator pays a cost, $1$, and gives its partner a benefit, $r$.
A defector incurs no cost and gives no benefit.
Thus, the payoff matrix can be expressed as follows.
\begin{align}
    \begin{pmatrix}
        r-1&-1\\
        r&0
    \end{pmatrix}
\end{align}
Since $0>-1$ and $r>r-1$, it is rational for every agent to choose to defect (i.e., the prisoner's dilemma).
Each node plays the prisoner's dilemma game with all its neighbors.
We can derive the AMEs for the model by replacing the payoff as the overall payoff $\Pi_{\mathrm{C}_m} = r m - k$, $\Pi_{\mathrm{D}_m} = r m$.
In this model, the points of discontinuity at $K=0$ are $r=k/n$ ($n \in \mathbb{N}$), 
and we can obtain
\begin{align}
    \frac{\d}{\d t} \rho^{\mathrm{C}} = - \frac{1}{2K} \sum_m (k-m) \rho^{*\mathrm{C}}_m < 0
\end{align}
at $K \gg 1$.
Thus, when the noise is sufficiently large, cooperators become extinct at any $r$.
Further, the lower boundary for the C and (C+D) phases must satisfy $r > k/(k-2)$.

Javarone et al.\cite{javarone2016conformity} have investigated the spatial public goods game in the presence of two types of agents.
Conformity-driven agents (CDAs) and fitness-driven agents (FDAs) update their strategies according to the transition rules of the voter model and Eq\eqref{imitation}, respectively.
Let $q_c$ ($q_f$) denote the fractions of CDAs (FDAs) ($q_c + q_f = 1$).
Assume that each agant does not change its behavior (i.e., conformity-driven and fitness-driven).
We can derive the AMEs for this model.
Let $\rho^{s}_{c,m}$ ($\rho^{s}_{f,m}$) be the fraction of $s_m$ nodes in CDAs (FDAs).
They satisfy the following relations:
\begin{align}
    \sum_{s\in \{\mathrm{C},\mathrm{D}\}}\sum_{m=0}^{k} \rho^s_{c,m} &= 
    \sum_{s\in \{\mathrm{C},\mathrm{D}\}}\sum_{m=0}^{k} \rho^s_{f,m} = 1,\\
    \rho^{s} &\equiv \sum_{x \in \{c,f\}} q_x \sum_{m=0}^{k} \rho^{s}_{x,m}.
\end{align}
We replace $p_{s}^{s'}(m')$ by the following:
\begin{align}
    p_{\mathrm{C}}^{\mathrm{C}}(m')
    &= \frac{\sum_{x} q_x m' \rho^{\mathrm{C}}_{x,m'}}{\sum_{x} q_x \sum_{m} m \rho^{\mathrm{C}}_{x,m}},&&
    p_{\mathrm{C}}^{\mathrm{D}}(m')
    = \frac{\sum_{x} q_x m' \rho^{\mathrm{D}}_{x,m'}}{\sum_{x} q_x \sum_{m} m \rho^{\mathrm{D}}_{x,m}},\nonumber\\
    p_{\mathrm{D}}^{\mathrm{C}}(m')
    &= \frac{\sum_{x} q_x (k - m') \rho^{\mathrm{C}}_{x,m'}}{\sum_{x} q_x \sum_{m} (k - m) \rho^{\mathrm{C}}_{x,m}},&&
    p_{\mathrm{D}}^{\mathrm{D}}(m')
    = \frac{\sum_{x} q_x (k - m') \rho^{\mathrm{D}}_{x,m'}}{\sum_{x} q_x \sum_{m} (k - m) \rho^{\mathrm{D}}_{x,m}}.
\end{align}
We define the transition probabilities as
\begin{align}
    W^{\mathrm{D} \to \mathrm{C}}_{c,m} \equiv W^{\mathrm{D} \to \mathrm{C}}_{m,\voter},&&
    W^{\mathrm{D} \to \mathrm{C}}_{f,m} \equiv W^{\mathrm{D} \to \mathrm{C}}_{m},&&
    \beta^{\mathrm{C}} = \sum_{x \in \{c,f\}} q_x  \sum_{m'} p_{\mathrm{C}}^{\mathrm{D}}(m') \ W^{\mathrm{D} \to \mathrm{C}}_{x,m'},&&
    \gamma^{\mathrm{C}} = \sum_{x \in \{c,f\}} q_x  \sum_{m'} p_{\mathrm{C}}^{\mathrm{C}}(m') \ W^{\mathrm{C} \to \mathrm{D}}_{x,m'}.
\end{align}
We define $\beta^{\mathrm{D}}$ and $\gamma^{\mathrm{D}}$ similarly.
Deriving Eq.\eqref{AME} for $\rho^{s}_{c,m}$ and $\rho^{s}_{f,m}$, respectively, gives the AMEs for this model.

\twocolumngrid

\end{document}